\documentclass[letterpaper,12pt]{article}
\usepackage{chicago,graphicx,amsmath}
\usepackage[margin=2cm,nohead]{geometry}
\usepackage[T1]{fontenc}
\usepackage[utopia]{mathdesign}

\title{Does Data Splitting Improve Prediction?}
\author{Julian J.\ Faraway\footnote{Department of 
Mathematical Sciences, University of Bath, BA2 7AY,
United Kingdom, \texttt{jjf23@bath.ac.uk}}}
\date{\today}

\begin{document}
\maketitle{}

\begin{abstract}
Data splitting divides data into two parts. One part is reserved for model selection. In some applications, the second part is used for model validation but we use this part for estimating the parameters of the chosen model. We focus on the problem of constructing reliable predictive distributions for future observed values. We judge the predictive performance using log scoring. We compare the full data strategy with the data splitting strategy for prediction. We show how the full data score can be decomposed into model selection, parameter estimation and data reuse costs. Data splitting is preferred when data reuse costs are high. We investigate the relative performance of the strategies in four simulation scenarios. We introduce a hybrid estimator called SAFE that uses one part for model selection but both parts for estimation. We discuss the choice to use a split data analysis versus a full data analysis.
\end{abstract}

\textbf{Keywords:} cross-validation, model assessment, model uncertainty, model validation, prediction, scoring.

\section{Introduction}
\label{sec:introduction}

Predictions based on statistical models are sometimes disappointing. We do not expect predictions to be exactly correct so we compute measures of uncertainty but the observed outcomes may lay well outside these expressions of the expected variation. Some failures cannot be anticipated as unexpected results can occur because the system generating the observables has changed since we collected the data used to develop the model. Alternatively perhaps our data was not broad enough to include more extreme events leading us to fail to anticipate such outcomes in our predictive model. However, sometimes the problem lies not in the data but in the statistical methods used.

In many applications, we do not know or choose to specify in advance the particular form of model to be used. We frequently use the data to select the model. Statisticians are aware of the problem of overfitting, namely using too complex a model resulting in a misleadingly good fit to the current data. We have become skilled at avoiding this problem by balancing fit against complexity. Nevertheless, once we have selected a model, we often proceed to estimate the parameters and make predictions with an assessment of variability that reflects the uncertainty about the parameters but not about the model. This lapse is the source of many statistician-generated failures with prediction.  See \shortciteN{Berk2009} and \shortciteN{Wit2012} for recent discussion of these issues and, less recently, in \citeN{Chatfield:95} and \citeN{potscher:91}.

Using the same data to both select and estimate the parameters has long been recognised as generating over-optimistic inference but practitioners frequently do little to address it. One reason for this lack of action is the difficulty in doing anything about the problem. One approach is to integrate the model selection and parameter estimation. For example, the Box-Cox method selects the index of transformation of the response in a regression problem. Instead, we might regard this as parameter estimation rather than model selection and proceed accordingly as in \citeN{hinkley:84}. Another example is the LASSO method which succeeds in combining regression variable selection with estimation although getting the subsequent inference correct is problematic. For more realistic complex model selection procedures that involve a combination of procedures and use both graphical and numerical approaches, a holistic approach to inference seems impractical. A Bayesian approach that assigns priors to models as well as parameters is possible as in \citeN{draper:95} but the approach becomes unworkable unless the space of models is small. In many cases, the space of models cannot be reasonably specified before analysing the data. Another idea is to using resampling methods to account for model uncertainty as in \citeN{faraway:92}. However, this method requires the model selection process to be pre-specified and automated. It also requires that these processes be implementable completely in software which excludes the possibility of human judgement in model selection.

There is a simple solution --- data splitting. We divide the data into two, not necessarily equal, parts. One part is used to select the model and the other part is used to estimate the parameters of the chosen model. The problem of  overconfidence in the chosen fit is avoided because we use fresh data to estimate the parameters of the model. Data splitting can easily be used in a wide range of problems and requires no special software. The analyst is free to select the model using any procedure which can involve subjective elements and need not be specified in advance. Unfortunately, there is a drawback. Less data is used to select the model so we cannot expect some reduction in the probability we select a better model. Furthermore, less data is used to estimate the parameters of the model so we can expect some additional uncertainty in prediction. Will the gain from avoiding data re-use compensate for the loss in model selection and parameter estimation performance? An answer to this question is the purpose of this paper.

\citeN{stone:74} provides an early history of data splitting. \citeN{Dawid1984} discusses the related but different strategy for updating models as new data arrive sequentially. In Section~\ref{sec:methods}, we discuss the uses of data splitting and the methods we use to evaluate its effectiveness. In Section~\ref{sec:simulations}, we present simulations to explore the relative value of data splitting. Section~\ref{sec:conclusion} contains our conclusions.

\section{Methods}
\label{sec:methods}

\subsection{Model validation}
\label{sec:model-validation}

Data splitting has been used for a variety of purposes. Having chosen and estimated a model, we may wish to evaluate how well the model can be expected to perform in practice. Under this setting, the first part of the data is used for model selection and estimation and second part is used to generate a measure of how well the model will predict future observations.  Data splitters \citeN{picard:84}, \citeN{picard:90} and \citeN{roecker:91} have an objective of estimating the average (over a subset of the predictor space) mean squared error of prediction. The aim for these authors is to obtain a better estimate of the quality of future predictions. Certainly, these methods will obtain more realistic estimates of this quality than using the naive estimate provided by the full data fit but this will come at a substantial price. Because less data is used for model selection and estimation, the prediction quality we are trying to estimate will be itself degraded.

In contrast to the selection-estimation data splitting, where we hope the splitting might result in better prediction, validation data splitting will almost surely make predictions worse. It depends how useful we find the estimate of prediction quality as to whether this loss is worthwhile. In some cases, analysts are in competition to produce the best predictive model. For example, in the Netflix prize competition (\citeN{Bell2007}) it was essential to hold back part of the data to evaluate the performance of the model since the competitors estimates of performance could not be trusted. Another example, is internet business Kaggle which challenges analysts to develop the best predictive method (\citeN{Carpenter2011}.  In other cases, there may be some regulatory requirement to assess model performance.

However, unless there is some reason not to entirely trust the analyst, there are better ways to estimate predictive performance than the naive method or validation data splitting. Simple methods like 10-fold cross-validation can be used which, while not perfect, will give more realistic estimates of future model prediction performance. Indeed, more recent authors discourage the use of data splitting for this purpose. \citeN{Steyerberg2009}, \shortciteN{Schumacher2007} and \shortciteN{Molinaro2005} all advise against data splitting although the latter two references investigate cases where the number of variables is large relative to the number of observations. In such cases one call ill-afford to give up observations for validation purposes.

This form of data splitting reserves part of the current data for validation purposes but the true test comes when genuinely new data arrive as \citeN{hirsch:91} points out. \citeN{Altman2000} also points out the distinction between statistical and clinical validity. All this is true but we shall restrict our concern to the data we have at hand and do the best we can with that. Some authors have proposed splitting data into three parts where one part is used for selection, the second for estimation and third for validation. These include \citeN[p.\ 13]{miller:90}, \citeN{mosteller:77} and \shortciteN{Friedman2008}. Some have used data splitting for quite different reasons --- see \shortciteN{Heller2009}. We have focussed on the problem of prediction but questions also arise in hypothesis testing --- see \citeN{Cox1975} and \shortciteN{Dahl2008}

\subsection{Scoring}
\label{sec:scoring}

We need a measure of quality for predictions. Predictive mean square error has been used frequently for this purpose but such a measure fails to judge the uncertainty expressed in the prediction. Confidence intervals for predictions can usually be generated and we can record whether the future observation falls within the specified confidence interval. But this is a crude measure that fails to use all the information. Suppose we specify a predictive distribution $f(y)$ and we subsequently observe $y_0$ then we assign a score of $-\log f(y_0)$. Log scoring was proposed by \citeN{good:52} and has been used commonly in fields such as weather forecasting where we must often judge the accuracy of forecasts. Other scoring rules have been proposed and discussed by \citeN{Gneiting2007} but log scoring appears most appropriate for this task as it is analogous to a log likelihood for new observations and the estimators we use in the simulations either are or are closely related to maximum likelihood estimators.  It is a proper scoring rule as defined by \shortciteN{Parry2012}.  

Log scoring is also used for binary response prediction where we use the probability rather than the density. This is also provides an intuitive way of interpreting a difference in scores. For example, a difference of 0.1 would represent a difference of about 10\% in predicting the correct result. Interpretation for continuous responses can be made using a difference in log-likelihoods with the discrete case providing some intuition for the interpreting the magnitude of these differences.

\subsection{Decomposition of Performance}
\label{sec:decomposition}

Consider a data generating process $Z=(X,Y)$.  We are given $n$ draws from $Z$ from which to construct a model that provides a predictive distribution for $Y$ given a value from $X$. We will subsequently be given new values from $X$ for which we are asked to construct predictive distributions for future values of $Y$. 
Suppose we entertain a set of models $\{M_i\}$ that provide for such predictive distributions. The set of models might be infinite in number and  might not contain the true data generating model. We have a model selection process $S$ such that given data of size $n$, $Z^j_n$, $S(Z^j_n) \rightarrow M_i$ indicating that given data $Z^j_n$, model $M_i$ is selected. 

Now suppose that after we have specified $S$ without knowledge of the truth about the data generating process, this information becomes available to us. We can then evaluate the performance of $S$ on samples of size $n$ from $Z$ by calculating:
\begin{displaymath}
  E_Z g(Z_n, M=S(Z_n))
\end{displaymath}
where $g(Z_n,S(Z_n))$ is the score for data $Z_n$ and model selected $S(Z_n)$ where a new $(X,Y)$ pair is drawn, a predictive distribution constructed and scored against the new value of $Y$. The expectation $E_Z$ is over both the data $Z_n$ and the future value used to evaluate the performance of $S$. Now set $S(Z_\infty) = M^*$ where $Z_\infty$ represents a  dataset of unlimited size enabling the best possible model $M^*$ to be chosen. Let $Z_n^F$ be a sample independent of $Z_n$. We can then decompose this measure of performance into four parts, respectively, best possible performance, model selection cost, parameter estimation cost and data re-use cost:
\begin{align}
  \label{eq:edec}
  E_Z g(Z_\infty, M^*) + E_Z I(S(Z_n) = M)[g(Z_\infty, M) - g(Z_\infty, M^*)] + \nonumber \\
  E_Z I(S(Z_n)=M)[g(Z^F_n, M) - g(Z_\infty, M)] + E_Z I(S(Z_n=M)[g(Z_n,M)-g(Z^F_n,M)]
\end{align}
In general, these components would be difficult to compute theoretically but are quite accessible by simulation. We can
draw $n_r$ samples of size $n$ from $Z$. To compute the scores, we will draw samples of size $n_e$ from $Z$ where $n_e >> n$. The performance can then be estimated using:
\begin{displaymath}
  {1 \over n_r} \sum_j g(Z^j, M_i = S(Z^j)) = {1 \over n_r} \left\{ \sum_{\mathrm{models}\; i} \sum_{j:S(Z^j)=M_i} g(Z^j, M_i) \right\}
\end{displaymath}
Let $p_i$ be the proportion of cases where model $M_i$ is selected, then we can estimate the four components using:
\begin{align}
  \label{eq:numdec}
  g(Z_\infty, M^*) + \sum_{Models\; i} p_i \left\{ g(Z_\infty, M_i) - g(Z_\infty,M^*) \right\} + \nonumber \\
\sum_{i} p_i \{g(Z^F,M_i) - g(Z_\infty, M_i)\}
+
{1 \over n_r} \sum_{j:S(Z_j)=M_i} \{ g(Z_j,M_i) - g(Z^F,M_i) \} 
\end{align}

The first term represents the best that can be done if we use the best model and we have an unlimited supply of training data denoted by the $Z_\infty$. We have no particular interest in this quantity but it is useful for scaling purposes.

The second term represents the model selection effect where more weight ($p_i$) put on sub-optimal models will increase the score (where high scores are bad). When a consistent model selection process is used, this term will tend to zero as the sample size increases. The difference between the full and split data values for this term will be of the same order as the rate at which $p_i$ tends to zero or one accordingly. The magnitude of the difference will depend also on the distribution of $g(Z_\infty,M_i)$ over ${M}$. The full data strategy will be favoured most when there is only one good model that is hard to find. Although the split data strategy will almost always trail the full data approach, the difference will not be large excepting situations with larger numbers of variables and smaller numbers of cases where the splitting of the data may cause the model selection to fail entirely.

The third term represents the cost of parameter estimation weighted by the choices of model. We have used $g(Z^F,M_i)$ to represent the empirically expected score under model $M_i$ where fresh estimation data is available. We could estimate this using only the replications where $M_i$ is selected but in the simulations to follow we can get a better estimate by computing it for all replications. The full data strategy will almost always produce lower scores for this component than the split data approach. The difference between the two will typically be $O(n^{-1/2})$. Excepting the situation where the split data is  insufficient to estimate the parameters, the difference between full and split is thus bounded.

The fourth term represents the cost of data reuse. For the split data approach, this term will have expectation zero because both $Z^j$  and $Z^F$ are divorced from the model selection process. For the full data strategy, it is difficult to say anything about the likely size of this term because of the complex relationship between model selection and estimation. As we shall see, this term can be large and can easily outweigh the advantages the full data had in selection and estimation. 

Thus the full data strategy will outperform the split data strategy on both model selection and parameter estimation but the difference is bounded and well understood because it is simply the effect of sample size. In contrast, the split data strategy will beat the full data strategy on data reuse cost. This cost could be very large. One must distinguish the problem of overfitting from that of model uncertainty. A statistician who tends to overfit will pay more in model selection and parameter estimation costs but
not necessarily more for data reuse --- consider the example of fitting too high an order in a polynomial regression. The statistician who wisely balances fit and complexity may reduce selection and estimation costs, but will still be exposed to data reuse costs when using a full data strategy.

\subsection{Estimators}
\label{sec:estimators}

We consider four estimators based on data $Z$ which is randomly split into a model selection set $Z_1$ and parameter estimation set $Z_2$
where the model selection set has size $[fn]$ for fraction $f$.
\begin{itemize}
\item FD: Full data estimate. $Z$ is used to both build/select the model and to estimate the parameters of that chosen model.
\item SD: Split data estimate. $Z_1$ is used to build/select the model and $Z_2$ is used to estimate the parameters.
\item SAFE: (Split Analysis, Full Estimate). $Z_1$ is used to build/select the model and $Z$ is used to estimate the parameters
\item VALID: Validation corrected estimate: $Z_1$ is used to build/select the model and estimate the parameters used to generate point predictions. $Z_2$ is used to generate a new estimate of the standard error to be used in the construction of predictive distributions.
\end{itemize}

The SAFE estimator is motivated by the decomposition of model performance. We see that the SD strategy will under-perform the FD approach for parameter estimation. But conditional on the model selection, the SAFE estimator will avoid this loss because all the data is used for estimation. On the other hand the FD strategy is vulnerable to heavy losses due to data re-use. Because the SAFE strategy withholds a portion of the data from model building, we have some protection against severe over-confidence because data not used in the model selection is used for the estimation. SAFE cannot mitigate any losses due to model selection.

The VALID estimator is motivated by the validation-motivated approach to data splitting. The second (or validation) sample may well reveal the overconfidence of our original predictions. We can form a new estimate of the standard error using 
\begin{displaymath}
  \hat\sigma^2 = \sum_i (y_i - \hat y_i)^2/([(1-f)n]-1)
\end{displaymath}
where the sum runs over $Z_2$. Such an approach is possible for location-scale type models like the Gaussian linear model but makes little sense for other types of model such as binary response regression. The VALID estimator will be effective for strategies that tend to overfit but it is difficult to see how it might affect data resuse costs.  The original predictive distributions might also be improved in other ways such as the one suggested by \citeN{Dawid1984}.

In \citeN{Little2006}, a suggestion is made that the model estimation should be Bayesian but the model assessment and checking should be Frequentist. This fits well with a data splitting approach because these activities can use different parts of the data.

\section{Simulations}
\label{sec:simulations}

It would be nice to explore the effects of data splitting mathematically. Unfortunately, this is very difficult. Finite sample calculations are required since asymptotically the issue becomes practically irrelevant. In \citeN{Leeb2005}, a calculation is made for the simple regression problem where the model selection is the determination of the significance of a single predictor. They demonstrate that problems with uniform convergence can arise for post-model selection estimators. Unfortunately, it becomes impractical to perform such calculations for richer and more realistic model selections scenarios. Hence, we resort to simulation.

\subsection{Box Cox}
\label{sec:box-cox}

The Box-Cox method selects the index of transformation on the response in a linear regression model. The simulation
setup is $X_i \sim U(0,1)$ and $\epsilon_i$  i.i.d. $N(0,\sigma^2)$ for 
$i=1, \dots n$ with
\begin{displaymath}
  Y_i^\lambda = \alpha + \beta X_i + \epsilon_i
\end{displaymath}
We fix $\alpha=0$ although the term will be estimated. 
We observe $X$ and $Y$ but not $\lambda$ (where $\lambda=0$ is equivalent to $\log Y$), which we estimate using the Box-Cox method. In keeping with common practice, we select $\lambda$ from a finite set of interpretable values, in this case \{-1.0, -0.5, 0.0, 0.5, 1.0\}. The $\alpha$ and $\beta$ are then estimated using least squares.  In this example, the model selection is just the determination of $\lambda$. 

We ran a simulation with $n_r=4000$ replications for each run using a full factorial design varying over all combinations of $\sigma = 0.1, 1, 10$, $\beta = 0, 1$, $n=18, 48$,  true $\lambda = -0.5, 0, 0.5$ and training fractions of $f=1/3, 1/2$ and $2/3$.

We computed the scores as follows: for each selected model, we generated 4000 realisations from the known true model which were used to score the fitted models (as we do in all subsequent simulations). We find the predictive distributions needed to compute the score using the predicted values and corresponding standard errors. There are more sophisticated ways to the construct frequentist predictive distributions, see for example \citeN{Lawless2005}, but we have consistently opted for the straightforward $t$-distribution scaled by the predicted mean and variance.

\begin{figure}
  \centering
  \includegraphics{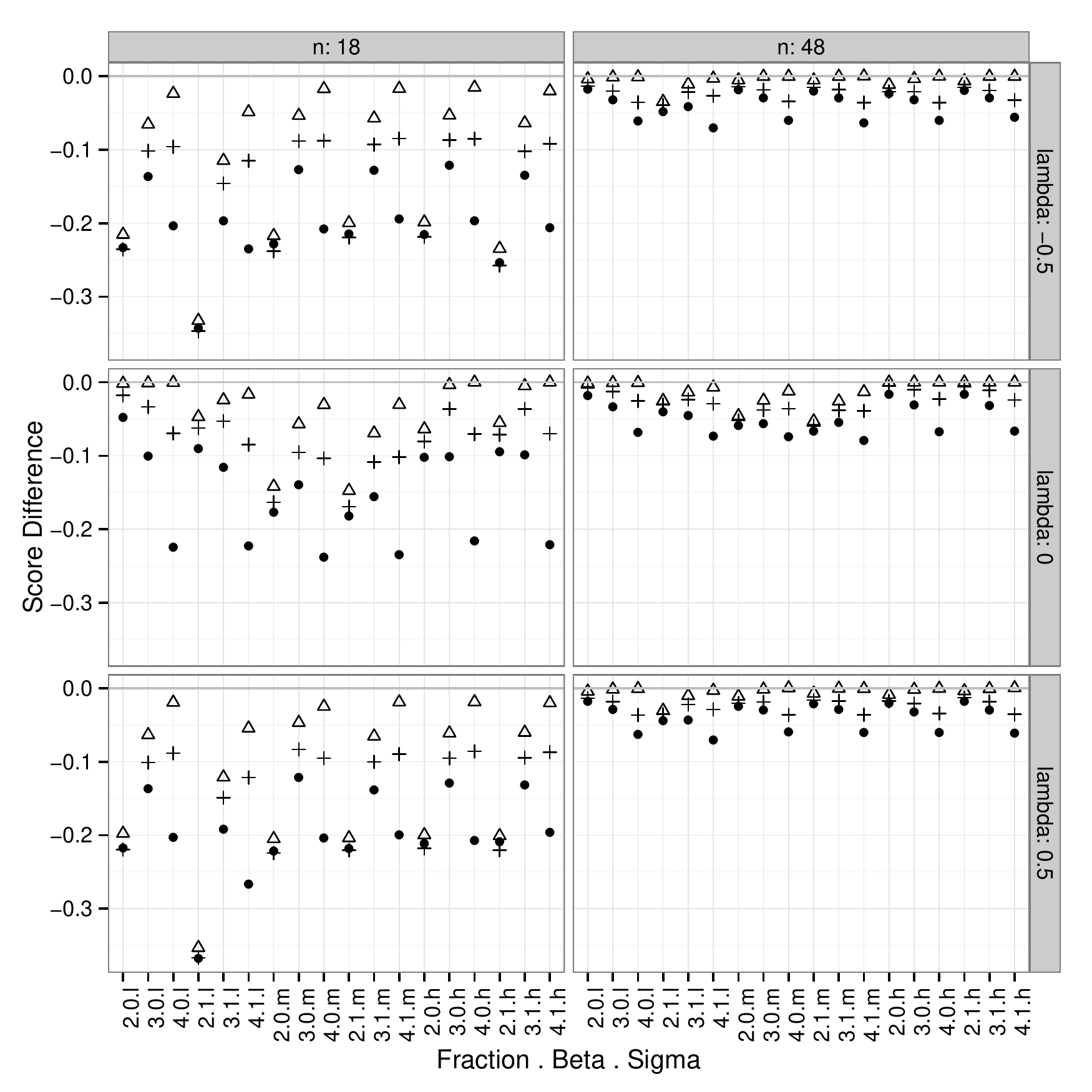}
  \caption{Box Cox simulation results showing the full minus the split score as a solid dot, full minus SAFE as an empty triangle and full minus valid as a plus symbol. The X-axis shows the variation over the training fraction (2/6, 3/6 or 4/6) and labelled as 2, 3 or 4, the value of $\beta=0,1$ and the value of $\sigma = 0.1, 1, 10$ denoted ``l'', ``m'' or ``h'' respectively. 
  \label{fig:bc}}
\end{figure}

In Figure~\ref{fig:bc}, we plot the difference in scores (FD-SD) , (FD-SAFE) and (FD-VALID). We see that FD is superior to SD in all combinations. The sample size has by far the largest effect on the margin of difference. For $n=48$, we see that the difference in score is a relatively small 3-5\%. All the other factors have little consistent effect on the difference on the score differences although a training fraction of 1/2 seems a good overall choice. The SAFE estimator is an improvement over the SD estimator in all cases and is often close in performance to the FD estimator, particularly if the 1/3 training fraction case is avoided. This might be expected as model selection cost would be largest in the $f=1/3$ case. The VALID estimator falls between the SAFE and SD estimators in performance.

\begin{figure}
  \centering
  \includegraphics{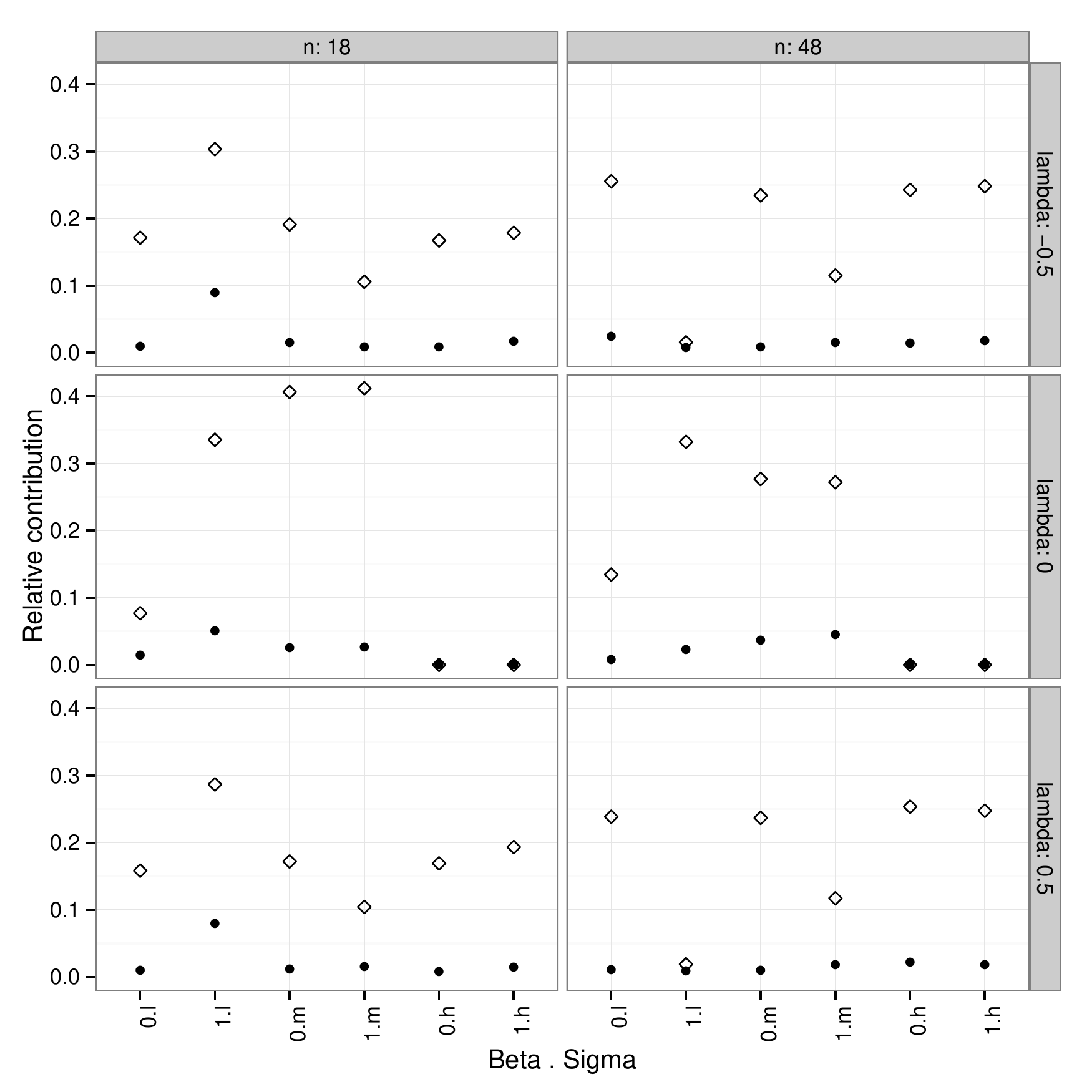}
  \caption{Model performance decomposition for the Box Cox scenario. Solid dot represents data reuse relative contribution while open diamond represent data reuse plus the model selection contribution. Parameter estimation is the remaining part completing the fraction to one.
  \label{fig:decbc}}
\end{figure}

Figure~\ref{fig:decbc} shows the estimated relative contribution as described in (\ref{eq:numdec}) for the full data case. We estimate the model selection, parameter estimation and data re-use costs and plot the proportion of the cost attributable to each part.  We average over the three training fraction runs at each level of the other simulation variables as this fraction has no impact on the full data estimator.  
Date reuse costs are minimal and parameter estimation costs account for a large share of the costs. Situations where the model selection costs are higher would be situations where the SAFE estimator might be expected to make the least improvement over SD --- there is some evidence from the plots to support this.

In the Box-Cox scenario, the full data strategy wins every time. There is very little model selection taking place with only five to choose from. With a small amount of data, the selection and estimation components favour the full data approach quite clearly and the data reuse costs are not significant. For the larger dataset, the difference between full and split was relatively small.

\subsection{Variable Selection}
\label{sec:variable-selections}

Consider a model:
\begin{displaymath}
  y = \alpha + \beta_1 X_1 + \beta_2 X_2 + \dots + \beta_p X_p + \epsilon
\end{displaymath}
where  $\epsilon_i$ is i.i.d. $N(0,\sigma^2)$, $X_i \sim U(0,1)$, $\mathrm{cor}(X_i,X_j) = \rho \quad \forall i \neq j$ and $\alpha=0$ but is estimated.
We follow a stepwise AIC-based variable selection procedure as described by the \texttt{step} function
of \citeN{venables02:_moder_applied_statis_s}.

\begin{figure}
  \centering
  \includegraphics{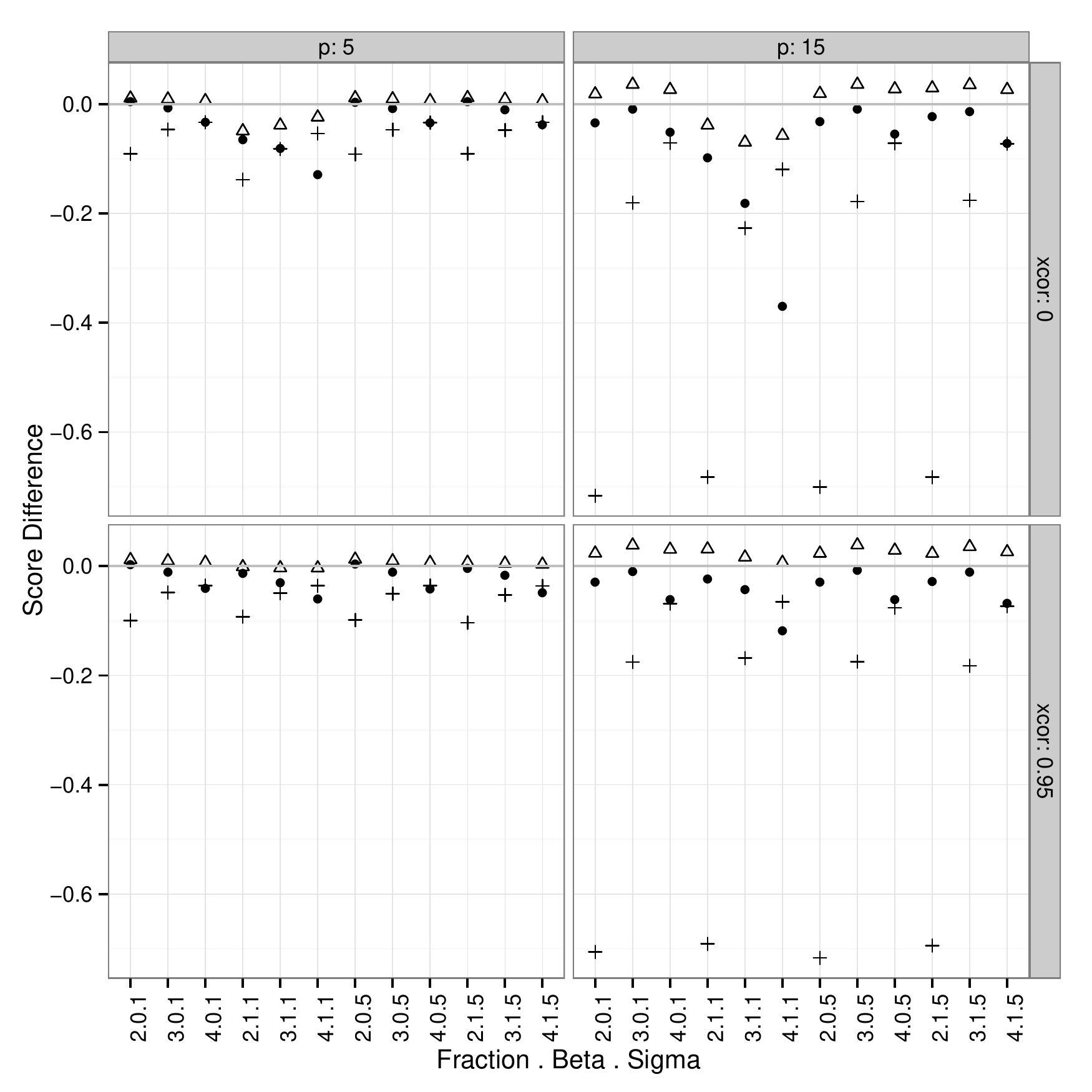}
  \caption{Variable selection simulation results showing the FD-SD difference in scores as a solid dot, FD-SAFE difference as an open diamond and FD minus VALID as a plus symbol.  X-axis shows the variation over the training fraction (2/6, 3/6 or 4/6) and labelled as 2, 3 or 4, the value of $\beta=0,1$ and the value of $\sigma = 1, 5$.
  \label{fig:vs}}
\end{figure}

The simulation used 4000 replications and $n=60$ with a full factorial design over all combinations of $\sigma = 1,5$, $\beta_i = 0,1 \forall i$, $p=5, 15$, $\rho=0, 0.95$ and training fractions of $1/3, 1/2$ and $2/3$. The results showing the average difference in score between the full and split analysis methods along with the FD-SAFE difference is shown in Figure~\ref{fig:vs}.

In Figure~\ref{fig:vs}, we see that FD is always better than SD but the margin of difference is mostly small. The best choice training fraction varies although f=2/3 is always worst. We see that the SAFE estimator is always better than the SD estimator and often outperforms the FD estimator. The VALID estimator can perform badly although it competes well with SD when $f=2/3$.

\begin{figure}
  \centering
  \includegraphics{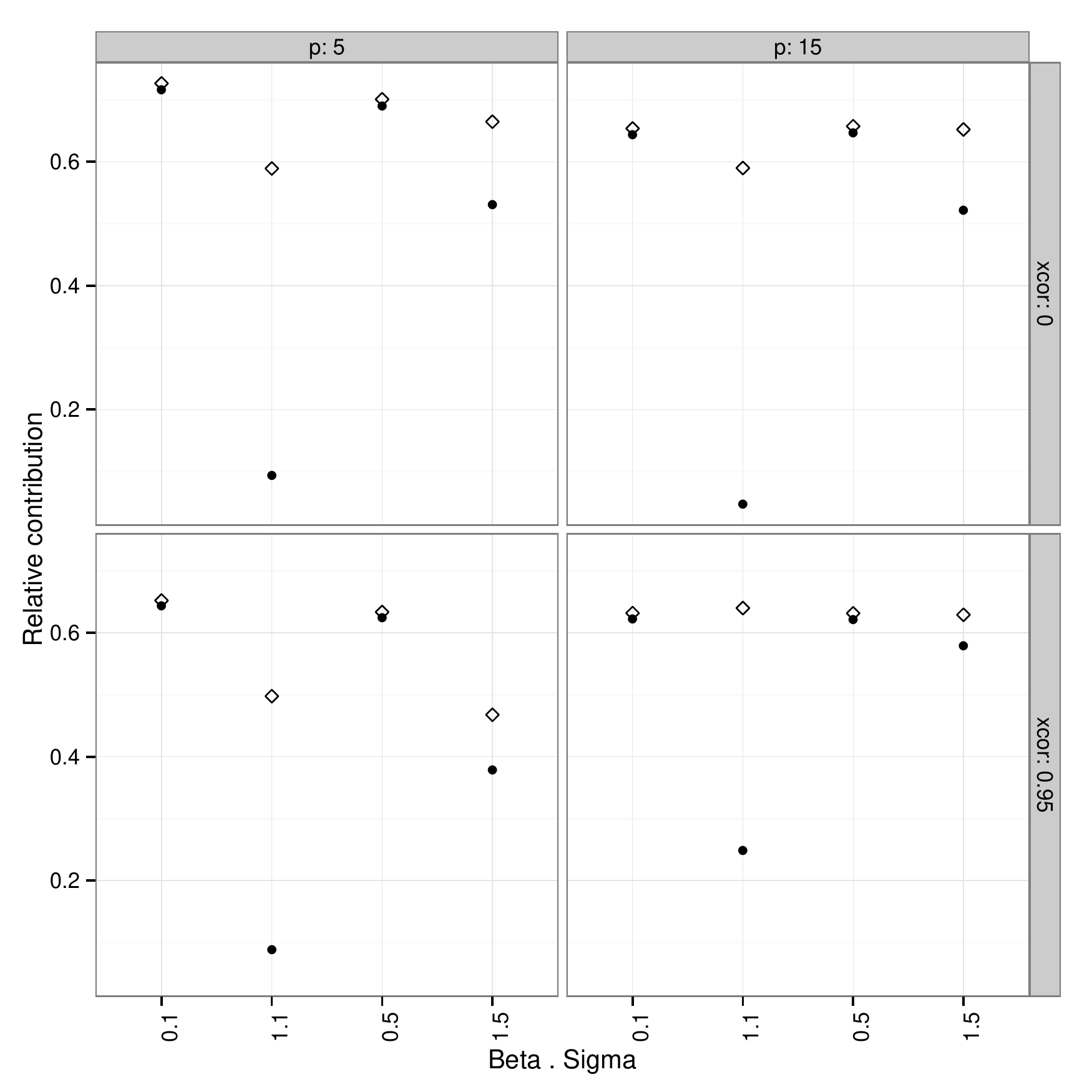}
  \caption{Model performance decomposition for the variable selection scenario. Solid dot represents data reuse relative contribution while open diamond represents data reuse plus the model selection contribution. Parameter estimation is the remaining part.}
  \label{fig:decvs}
\end{figure}

Figure~\ref{fig:decvs} shows the relative contribution as described in (\ref{eq:numdec}).  This shows that the data reuse costs can be quite significant which allows for split strategies to be favoured. The higher model selection costs for the $\beta=1, \sigma=1$ combination are reflected in the reduced performance of the split strategies as seen in Figure~\ref{fig:vs}.

We extended the data analysis method to allow consideration of potential second order terms in the predictors (even though these are not present in the true model). Under these conditions, the SD strategy tends to outperform FD.

\subsection{Outliers}
\label{sec:outliers}

In this scenario, we investigate a model building strategy that eliminates outliers. Consider a model with $X_i \sim U(0,1)$ with $i=1, \dots n$. Let $\epsilon_i$ be i.i.d. $t$ with degrees of freedom $d$.
Let
\begin{displaymath}
  Y_i = \alpha + \beta X_i + \epsilon_i
\end{displaymath}
where the true value of $\alpha=0$ although the model will estimate it. The model building strategy fits the model using least squares and computes the studentized residuals, $r_i$. Any case with $|r_i| > 3$ is deleted and the model is refitted. This case deletion process is repeated until all remaining $|r_i| < 3$. This is a crude procedure and one which we would not recommend, particularly when we can see the true model generating process. Nevertheless, it is representative of procedures that delete or down-weight cases that do not fit the proposed model well. We would prefer an approach to prediction to behave reasonably well even if the model building strategy is not the best.

\begin{figure}
  \centering
  \includegraphics{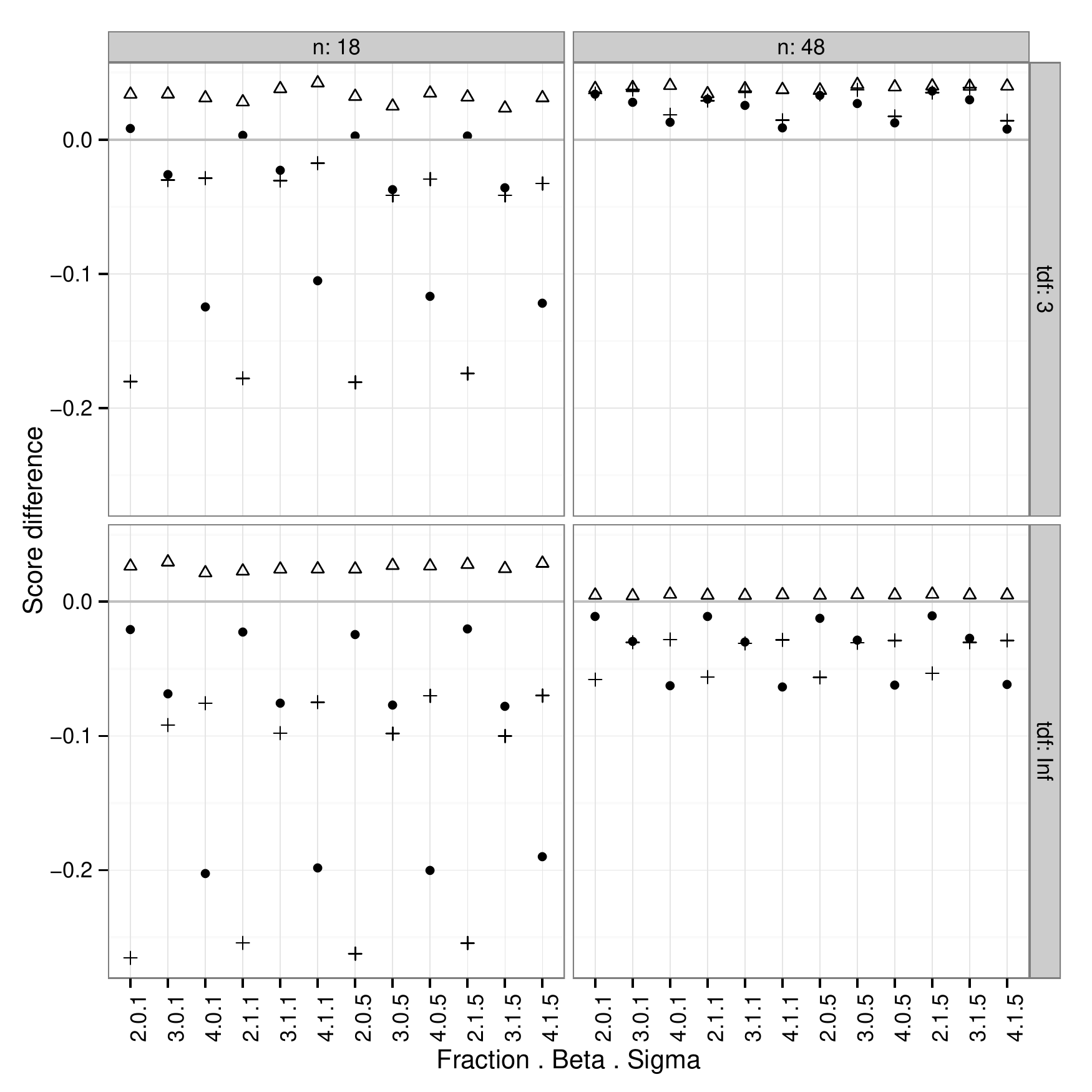}
  \caption{Outlier simulation results showing the full minus the split score as a solid dot, FD-SAFE as a diamond and FD-VALID as a plus symbol. . X-axis shows the variation over the training fraction (2/6, 3/6 or 4/6) and labelled as 2, 3 or 4, the value of $\beta=0,1$ and the value of $\sigma = 1, 5$.
  \label{fig:outlier}}
\end{figure}

We ran a simulation with 4,000 replications. We used a full factorial design running over all combinations of $n=18,48$,
$\sigma = 1,5$, $\beta = 0,1$, $d=3, \infty$ and training fractions of $1/3, 1/2$ and $2/3$. We see the outcome in
Figure~\ref{fig:outlier}.

When the error is $t_3$ we see that the SD approach is superior for the larger sample but inferior when the error is $t_\infty$ i.e. normal.
In all cases, the smallest training fraction is preferred for SD but this is simply because the model selection process here just decides which points to include. This choice becomes irrelevant because the test data is used to estimate the model.
The other factors have little effect on the outcome. We see that the SAFE estimator performs very well in this instance although the superiority over the full data strategy is due to the sensible decision to use all the data to estimate the model. The VALID estimators performs similarly to the SD estimator except that the preferred $f=2/3$.
The outlier deletion strategy is clearly artificial in this scenario but this does show how this type of data analytic procedure,
which is often used in more plausible forms in practice, can have serious consequences for prediction performance

There is no model selection effect in this scenario as the model is fixed, only the data used vary. We can compute the decomposition in (\ref{eq:numdec}) and find that the data reuse costs dominate the parameter estimation comprising a large proportion the total contribution of these two components across the 48 simulation conditions (plot not shown).

\subsection{Binary Response}
\label{sec:binary-response}

We finish with a variable selection problem in a binary response logistic regression model.
Let $\eta_i = \sum_{j=1}^p \beta_j X_{ij} + \epsilon_i$ for $i=1,\dots,n$ where $\epsilon_i$ is i.i.d. $N(0,\sigma^2)$ and $X_{ij} \sim U(0,1)$.
Now let $P(Y_i = 1) = \exp(\eta_i) / (1 +\exp(\eta_i))$.

The model building strategy is variable selection using stepwise AIC as implemented in \citeN{venables02:_moder_applied_statis_s}. We vary $p=1, 3, 5$, $n=18, 48$, $\sigma = 0.1, 1$, $\beta=0, 1$ and 
training fractions of $1/3, 1/2$ and $2/3$. We used 4000 simulation replicates each over all factorial combinations of these 5 parameters.

\begin{figure}[hbtp]
  \centering
  \includegraphics{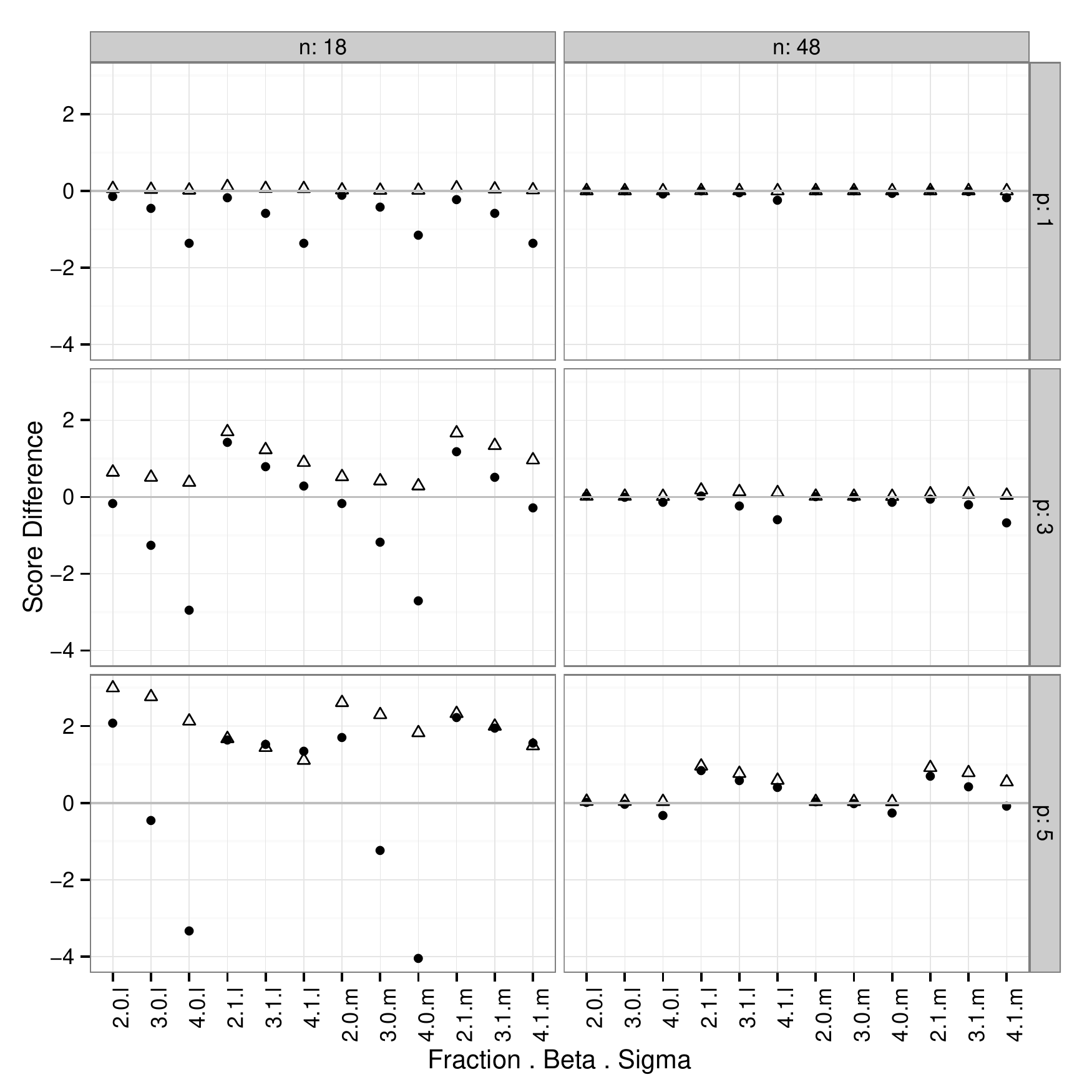}
  \caption{Binary response simulation results showing the full minus the split score. X-axis shows the variation over the training fraction (2/6, 3/6 or 4/6) and labelled as 2, 3 or 4, the value of $\beta=0,1$ and the value of $\sigma = 0.1, 1$ denoted ``l'' or ``m''  respectively. 
  \label{fig:bin}}
\end{figure}

The results are shown in Figure~\ref{fig:bin}. We see that the results are much more stable for n=48 than for n=18. For the simplest $p=1$ where there is very little model selection, the full data strategy is preferable. When there are more predictors, the split data strategy is preferred when there really are significant predictors ($\beta=1$) but the full data approach works best when the null model is true. We see that the SAFE estimator is almost always an improvement over the SD estimator and sometimes better than the FD estimator depending on the scenario. Note the the VALID estimator makes no sense for a binary response model.

\section{Conclusion}
\label{sec:conclusion}

Our main conclusion is that, except in specific circumstances, a data splitting strategy is superior to a full data strategy for prediction purposes.  Caution is necessary when trying to generalise from simulations which are necessarily limited in scope. But we must bear in mind that finite sample theoretical results of any generality are not available. Prediction from statistical models is a common activity and practical advice is necessary.

The simulation results are not unfavourable to the FD strategy but the simulations understate the relative value of an SD strategy. Simulations cannot simulate human judgement which is an important part of model building which implicitly increases the number of potential models substantially. Increasing model choice favours an SD strategy since data reuse costs are amplified. Additionally,  all four scenarios we have presented are relatively simple which favours the FD strategy. Furthermore, the model building strategy has been completely specified. Under such conditions we would tend to choose an FD strategy but even here, we see that the SD and particularly the SAFE estimator might be a better choice. In many real data analyses on observed, not simulated data, the model building strategy will be more complex and difficult or impossible to pre-specify. In such circumstances, a split data strategy is preferable. The decomposition of the model performance indicates that the loss in using a split data approach is limited for model selection and parameter estimation once we move away from the $p$ close to $n$ situations. On the other hand, the full data approach is susceptible to unlimited losses due to data re-use costs. This suggests that the split strategy is safer than the full data strategy.

We make the following tentative recommendations --- in regression problems a continuous or binary response with independent observations (i.e. having no grouped, hierarchical or serial structure), the analyst should decide between a full and a split analysis based on the following considerations.

A full data analysis should be considered where models considered involve a number of parameters approaching half the number of observations or more. A full analysis is also preferred when the model building step will be simple and involve a choice among a small set of possibilities. In situations where the analyst is prepared to pre-specify the exact form of the data analysis in advance, bootstrapping or other resampling techniques can be used to account for model uncertainty. A full data analysis is likely to be preferable in designed experiments where there is often limited model choice.

A split data analysis should be preferred when the analyst has no fixed conceptions about the model to be used for the data and plans to search for a good choice using a range of numerical and graphical methods. The SAFE estimate should be preferred to the SD estimate. If it becomes clear that there is insufficient data in the split datasets to find and estimate a model, then the analyst can always revert to the full data approach (but a switch from full to split cannot be allowed). The empirical evidence gives no clear choice about the data splitting fraction but a 50:50 split seems a good default choice.

More work is necessary before attempting recommendations for other types of response or data of a more structured form. It is also more difficult to make a recommendation for situations where an explanation of the relationship between $X$ and $Y$ is the goal rather than prediction. Certainly the same issues of overconfidence arise but interpretation of parameters explaining the relationship become problematic when the form of the model is changing.

\bibliography{jfpapers}
\bibliographystyle{chicago}

\end{document}